# Strong polarization of individual nuclear spins weakly coupled to nitrogen-vacancy color centers in diamond

*Jiwon Yun, Kiho Kim, and Dohun Kim\**

*Department of Physics and Astronomy, and Institute of Applied Physics, Seoul National University, Seoul 08826, Korea*

*\*Corresponding author: dohunkim@snu.ac.kr*

**Abstract**

We experimentally demonstrate high degree of polarization of $^{13}$C nuclear spins weakly interacting with nitrogen-vacancy (NV) centers in diamond. We combine coherent microwave excitation pulses with optical illumination to provide controlled relaxation and achieve a polarity-tunable, fast nuclear polarization of degree higher than 85% at room temperature for remote $^{13}$C nuclear spins exhibiting hyperfine interaction strength with NV centers of the order of 600 kHz. We show with the aid of numerical simulation that the anisotropic hyperfine tensor components naturally provide a route to control spin mixing parameter so that highly efficient nuclear polarization is enabled through careful tuning of nuclear quantization axis by external magnetic field. We further discuss spin dynamics and wide applicability of this method to various target $^{13}$C nuclear spins around the NV center electron spin. The proposed control method demonstrates an efficient and versatile route to realize, for example, high-fidelity spin register initialization and quantum metrology using nuclear spin resources in solids.

**Introduction**

Nuclear spins, owing to their extremely long coherence times, have emerged as attractive candidates for solid-state quantum information processing and quantum-sensing applications [1-4]. Among the various physical platforms, nuclear spins interacting with nitrogen-vacancy (NV) center electron spins in diamond are of great interest as this spin system provides an unique method for measuring nuclear spin state through the measurement of NV centers' well-known spin dependent fluorescence and optically detected magnetic resonance [5].

Quantum manipulation using magnetic resonance plays a central role in this field of research [6]; and multi-spin-state initialization is a crucial step to achieve high-fidelity quantum manipulation such as entanglement generation [7] and error correction [8]. For the nuclear spins in diamond, an optically induced dynamic nuclear polarization method at the excited [9] or ground-state level anti-crossing [10] is conventionally used for high-fidelity polarization of the nitrogen nuclear spin inherent to the NV centers. On the other hand, for the spatially distributed $^{13}$C nuclear spins, the random magnitude of hyperfine tensor components and nuclear quantization axis differing from NV axis typically lead to much less polarization efficiency even at level anti-crossings [11]. Several strategies have been developed with widely varying degree of polarization and applicable external magnetic field range. Representative examples include dynamical-decoupling-based pulsed polarization [12], Hartmann–Hahn type cross polarization [13-15], and double resonance and double-π-pulse-based population transfer [16] method. Most of the existing polarization method, however, are targeted either for moderately or strongly coupled nuclear spins to the NV centers (hyperfine interaction > 2 MHz) [11], often require long polarization sequence duration [16,17], or targeted for large external magnetic field [9]. Since the hyperfine coupling strength of remote nuclear spins decreases rapidly as the distance between the two spins increases, there are limitations in expanding the spin system of the NV center via nuclear spins. By developing additional polarization methods to polarize even the weakly coupled nuclear spins, the size of the nuclear spin system accessible for quantum information or metrology can be expanded. Therefore, it is important to develop novel methods for efficient and fast polarization of weakly coupled nuclear spins in diamond.

In this study, we have demonstrated a novel $^{13}$C nuclear spin polarization method in diamond using energy-level selective microwave pump pulse and optical illumination. We exploited a tunable nuclear spin mixing parameter through the external magnetic field and showed that the interplay between the microwave pump rate and the optically induced relaxation rate leads to a population transfer-based nuclear polarization, which is higher than 85% at room temperature even for remote $^{13}$C nuclear spins exhibiting hyperfine interaction with NV centers of the order of 600 kHz. Moreover, we show that the method naturally enables selectivity on the target nuclear spin as well as controllable polarity with respect to the NV axis by adjusting the pump pulse frequency. A numerical simulation was performed using experimentally reasonable parameters to show the detailed spin dynamics and its dependence on the physical parameters such as hyperfine interaction dependent external magnetic field and polar coordination of $^{13}$C. With a room for further improvement and applicability to wide range of nuclear spins near NV center, we expect the present control method of nuclear-spin states can be used for high fidelity initialization of spin registers.

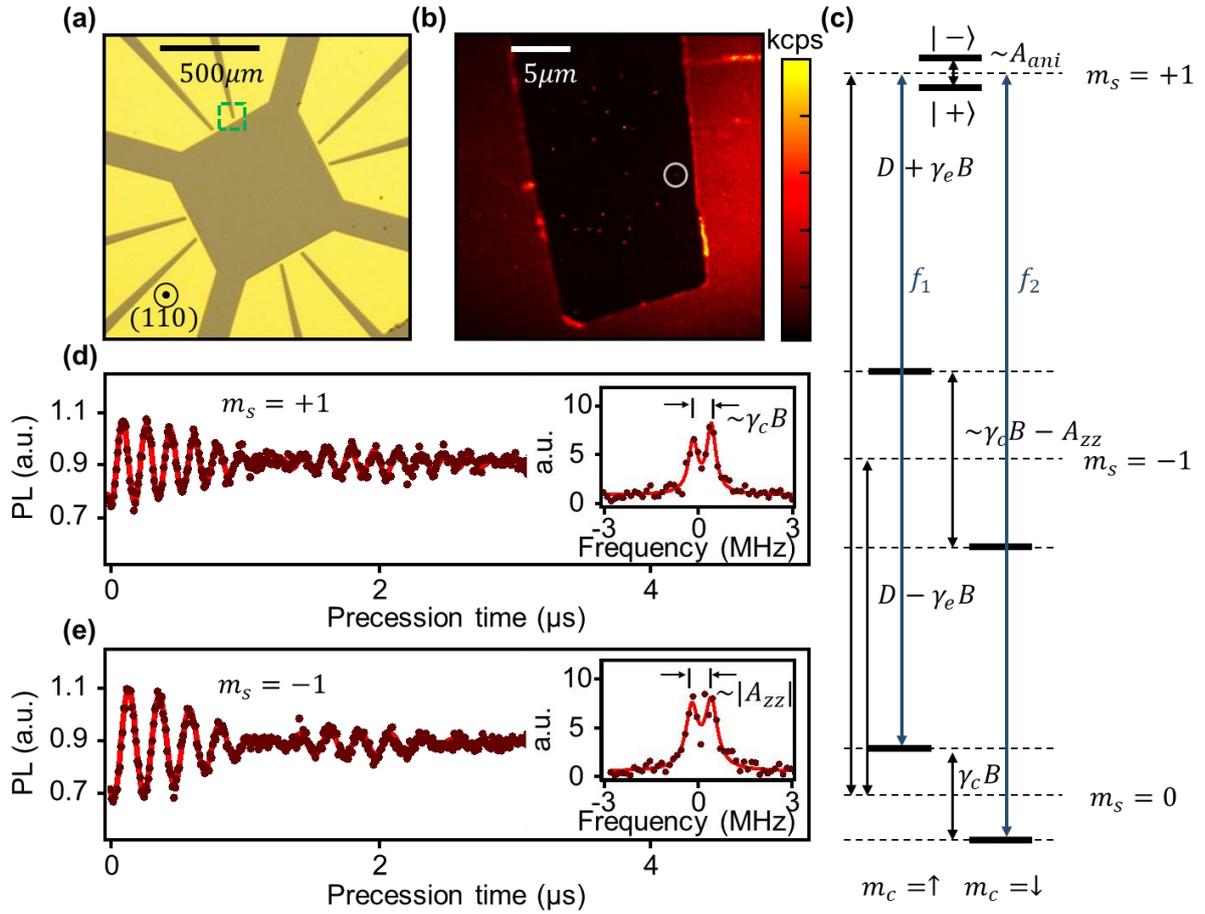

**Figure 1. Eigen energy space and coherence properties of the NV center. a)** A bright field image of the surface of the diamond sample with the fabricated coplanar waveguide. **b)** 2D confocal scan image of the area depicted by the green square in **a)**. The gray circle indicates the specific NV center that was investigated in this study. The color scale represents the photon count per second (cps) measured for each point on the 2D image scaled from 0 to 50 kcps. **c)** A schematic of the energy-level structure of the NV center interacting with $^{13}C$ under the aligned external magnetic field; $f_1$ (or $f_2$) represents the frequency corresponding to the energy-level difference between the $m_s$ = 0 and +1 state when $m_c$ = ↑ (or ↓). **d,e)** Result of the Ramsey oscillation at: **d)** $m_s$ = +1 and **e)** $m_s$ = -1. The inset graph shows the fast Fourier transform (FFT) of each Ramsey oscillation. The splitting of the two peaks in the FFT

represents: **d)** $\gamma_c B_z$ and **e)** $A_{zz}$. The uneven size of the two peaks of the FFT result comes from the following features: the low frequency resolution of the FFT result inducing fitting errors and a weak polarization originating from the spin mixing in the excited state. The red solid curves are fits to superposition of two sinusoidal functions with Gaussian decaying envelop.

**Methods**

We studied a CVD-grown high-purity type-IIa bulk diamond with nitrogen concentration < 10 ppb and natural abundance of $^{13}$C (~1.1%), as shown in Fig. 1a. The NV centers investigated in this study are the naturally formed color centers located roughly 2 μm below the {110}-oriented surface. We used a home-built confocal microscope to measure the optically detected magnetic resonance [18] and microfabricated a coplanar waveguide on the surface of the bulk diamond for applying the control pulses as shown in Fig. 1a. Figure 1b shows the magnified confocal microscope image revealing the spatially resolved single NV centers near one arm of the waveguide. An external magnetic field $B_z$ of ~520 G was applied along the direction to the studied NV center axis with an alignment precision better than 0.1° [19]. At this field, the $^{14}$N nuclear spin intrinsic to the NV center is polarized to its magnetic quantum number $m_n = +1$ state with a degree higher than 98% through an optically induced nuclear spin mixing mechanism, which can be obtained by the excitation laser illumination during the readout process of each measurement sequence [9]. Thus, we do not consider the nitrogen nuclear spin degree of freedom in the present study. We used time-domain Ramsey spectroscopy to identify $^{13}$C nuclear spins weakly coupled to the NV center electron spin with the magnitude of hyperfine interaction parallel to the NV axis $|A_{zz}| < 1$ MHz.

Figure 1c depicts the spin system considered in this study. In this two-spin system, the six eigen energy states are represented in terms of the superposition of the basis states of the electron spin quantum number $m_s$ = 0, -1, 1 and the $^{13}$C nuclear spin quantum number $m_c$ = ↑,↓ , because the anisotropic hyperfine tensor component $A_{ani}$ combined with external magnetic field leads to nonzero nuclear spin mixing. In particular, for $m_s$ = +1 manifold the nuclear spin eigenstates are given by the following equations:

$$|+\rangle = \cos(\theta/2)|\uparrow\rangle + \sin(\theta/2)e^{i\phi}|\downarrow\rangle \qquad (1)$$

$$|-\rangle = -\sin(\theta/2)e^{-i\phi}|\uparrow\rangle + \cos(\theta/2)|\downarrow\rangle \qquad (2)$$

Here, the nuclear mixing parameter $\theta$ can be calculated as $\tan\theta = A_{ani}/(A_{zz} + \gamma_c B_z)$, which is adjustable with the external magnetic field, where $\gamma_c \approx 1.07$ kHz/G is the nuclear gyromagnetic ratio of $^{13}$C and $\phi$ is a phase set by the hyperfine tensor components (see Appendix B for more details). In this experiment, we focused on a $^{13}$C nuclear spin that is weakly coupled to the NV center with $|A_{zz}| \sim 600$ kHz $\gg A_{ani}$, and $A_{zz} < 0$. With the application of $B_z \sim 520$ G along the NV axis, we work in the regime where $A_{zz} + \gamma_c B_z \sim 0$; thus $\theta \sim \pi/2$ so that the nuclear spin flipping transitions are equally allowed in the $m_s$ = +1 manifold. On the other hand, for $m_s$ = -1, the manifold mixing parameter $\theta'$ is set by $\tan\theta' = A_{ani}/(-A_{zz} + \gamma_c B_z) \sim 0$ and the nuclear spin mixing is minimized. We point out that only the mixing parameter $\theta$ is involved in this polarization method, and it does not show explicit dependence on $|A_{zz}|$ itself.

Figures 1d and 1e show the coherence property of the NV electron spin observed by Ramsey spectroscopy at $B_z$ = 520 G for $m_s$ = +1 (Fig. 1d) and $m_s$ = -1 (Fig. 1e). With the inhomogeneous coherence time $T_2^*$ ~ 2 μs, the Ramsey oscillations show beating due to the coexisting probability of the $^{13}$C nuclear spin initialized to $m_c$ = ↑ or ↓ states. The beating has a frequency splitting of ~600 kHz in both Fig. 1d and 1e (see fast Fourier transform (FFT) data as inset to Fig. 1d and 1e, indicating $\gamma_c B_z$ ~ 600 kHz (from Fig. 1d) as well as $|A_{zz}|$ ~ 600 kHz (from Fig. 1e); also see Appendix C for more details). The FFT spectrum shows nearly equal amplitude for the two frequency components indicating that the initial nuclear spin is in a completely mixed state.

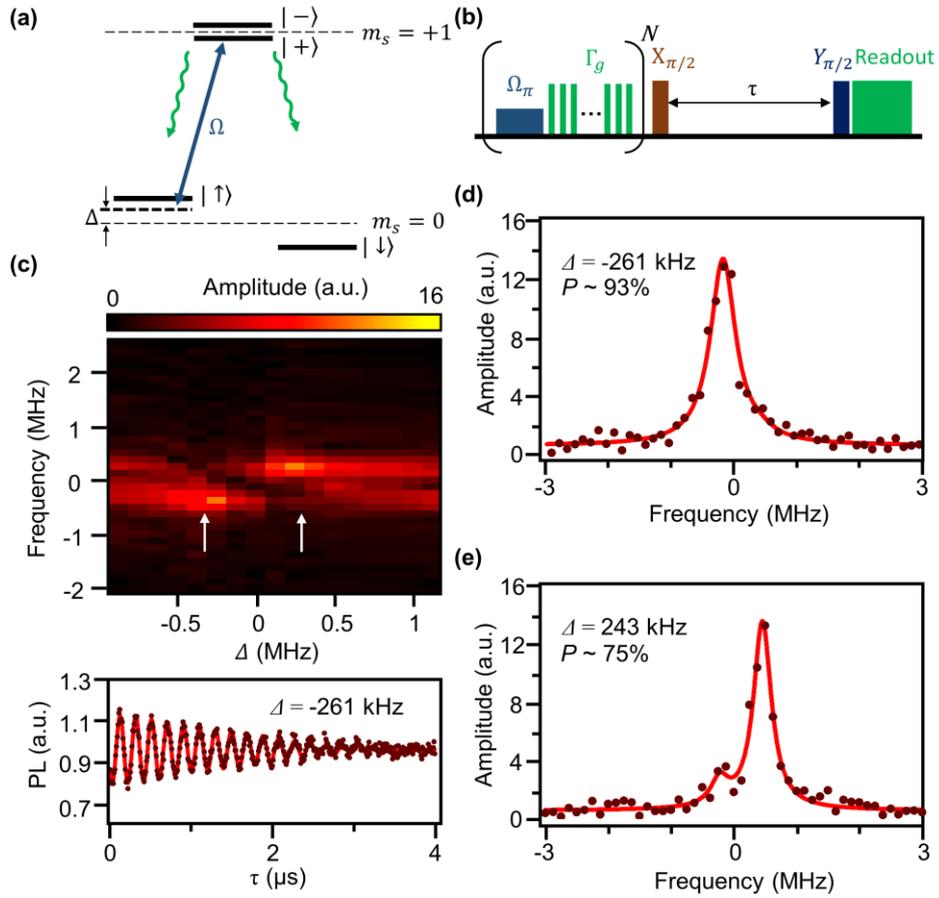

**Figure 2. Efficient nuclear polarization of weakly coupled $^{13}$C spin. a)** Subspace of the spin system considering $m_s = 0$ and $m_s = +1$ used for nuclear polarization. The green (or blue) arrows show the role of the green laser (or microwave) applied to this system; $\Delta$ is the detuning of the microwave with a power corresponding to the frequency of the Rabi oscillation $\Omega$. **b)** The pulse sequence used for the polarization of the $^{13}$C nuclear spin and the subsequent Ramsey measurement. **c)** FFT of the Ramsey oscillation measurement result while sweeping the detuning of the polarization microwave. The two lines with large amplitude represent the two different oscillating frequencies coming from different energy levels. The two white arrows

indicate the point from which the maximum polarization curves were extracted. The bottom panel is an example of the time-domain result of the Ramsey sequence near the maximum nuclear polarization. Red solid curve is a fit to Gaussian decaying sinusoidal function. **d,e)** Ramsey spectrum showing experimentally observed maximum nuclear spin polarization to: **d)** $m_c = \downarrow$ and **e)** $m_c = \uparrow$ states. The red solid curves are fit to Lorentzian functions.

We now discuss the mechanism of the proposed nuclear spin polarization scheme focusing on the $m_s = 0$ and $+1$ manifolds: We consider an initial completely mixed state composed of $|m_s, m_c\rangle = |0, \uparrow\rangle$ and $|0, \downarrow\rangle$ and applied selective microwave π pulse on this spin system, providing an initial nuclear spin dependent excitation. As mentioned earlier, we work in the $\theta \sim \pi/2$ regime, where the microwave π pulse excites the initial state to both nuclear spin projections, leading to efficient spin mixing at the $m_s = +1$ manifold, which is subsequently relaxed back to the $m_s = 0$ manifold by optical illumination, as shown in Fig. 2a. Because of this unbalanced pumping and relaxation path, the system is polarized to some degree to one of the nuclear spin states after the repeated application of the sequence. We note that a similar eigenspace of the coupled NV electron and $^{13}$C nuclear spin system, along with double resonant microwaves, has been used to demonstrate coherent population trapping in microwave regime in diamond, preparing the system state to the dark state [20]. Likewise, we consider the proposed nuclear spin polarization method as a preparing the system in the dark state using a single resonant microwave driving starting from the statistical mixture.

Generally, the limiting factor for nuclear spin polarization is the nuclear spin mixing that occurs at optically excited levels, exhibiting significant nuclear spin mixing induced by hyperfine interaction [21]. To minimize this unwanted additional spin mixing, we used a chopped laser pulse train with a pulse length of approximately 30 ns and a duty cycle ~30%, where a single-pulse duration is shorter than the time scale for a typical nuclear spin mixing; however, repeated application ensures electron spin relaxation as theoretically and experimentally considered in recent studies [21]. After applying the polarization sequence with a given repetition number $N$, we measured the degree of polarization using Ramsey spectroscopy. The whole experimental pulse sequence including microwave and laser is shown in Fig. 2b.

**Results and discussion**

We performed Ramsey spectroscopy while sweeping the microwave frequency used in the π pulse of the polarization sequence, resulting the 2D plot shown in Fig. 2c. The bottom panel of Fig. 2c is an example of a time domain result of the Ramsey sequence near the resonant condition showing only a single frequency component while maintaining the full optical contrast. Figures 2d and 2e show the results of a representative polarization when the microwave pump frequency $f_{mw}$ is resonant with the transition frequencies $f_1$ (d) and $f_2$ (e), respectively (see Fig. 1c for definitions of $f_1$ and $f_2$). In this experiment, we used $N = 6$ and Rabi amplitude $\Omega$ of the microwave pump of ~300 kHz.

Considering the polarization of $^{13}C$ either to $|\uparrow\rangle$ or $|\downarrow\rangle$, we quantified the degree of

polarization as follows:

$$P = \frac{\mathcal{P}_\uparrow - \mathcal{P}_\downarrow}{\mathcal{P}_\uparrow + \mathcal{P}_\downarrow} \quad (3)$$

where $\mathcal{P}_\uparrow$ (or $\mathcal{P}_\downarrow$) is the population of the $|\uparrow\rangle$ ($|\downarrow\rangle$) state of the $^{13}$C nuclear spin with respect to the NV axis, which we extracted from the experiment by fitting the FFT of the Ramsey data to two Lorentzian functions with different center frequencies. We observed that the value of $P$ approaches ~75% (93%) to the $|\uparrow\rangle$ ($|\downarrow\rangle$) orientation at the corresponding resonant conditions.

The asymmetric maximum polarization values are likely because the experimental microwave sweep resolution was low in the $|\uparrow\rangle$ orientation range, and the error of fitting the FFT result cannot be ignored because of the lack of points used for fitting; however, we show below that generally symmetric maximum polarization to the $|\uparrow\rangle$ ($|\downarrow\rangle$) orientation is expected theoretically. Moreover, unlike previous studies [16,17], our method does not require a slow rf nuclear Rabi pulse because the nuclear spin mixing is naturally allowed in the $\theta \sim \pi/2$ regime tuned by the external magnetic field. The results thus show a promising method for realizing a simple, target selectable, and high degree of polarization for weakly interacting nuclear spins in diamond.

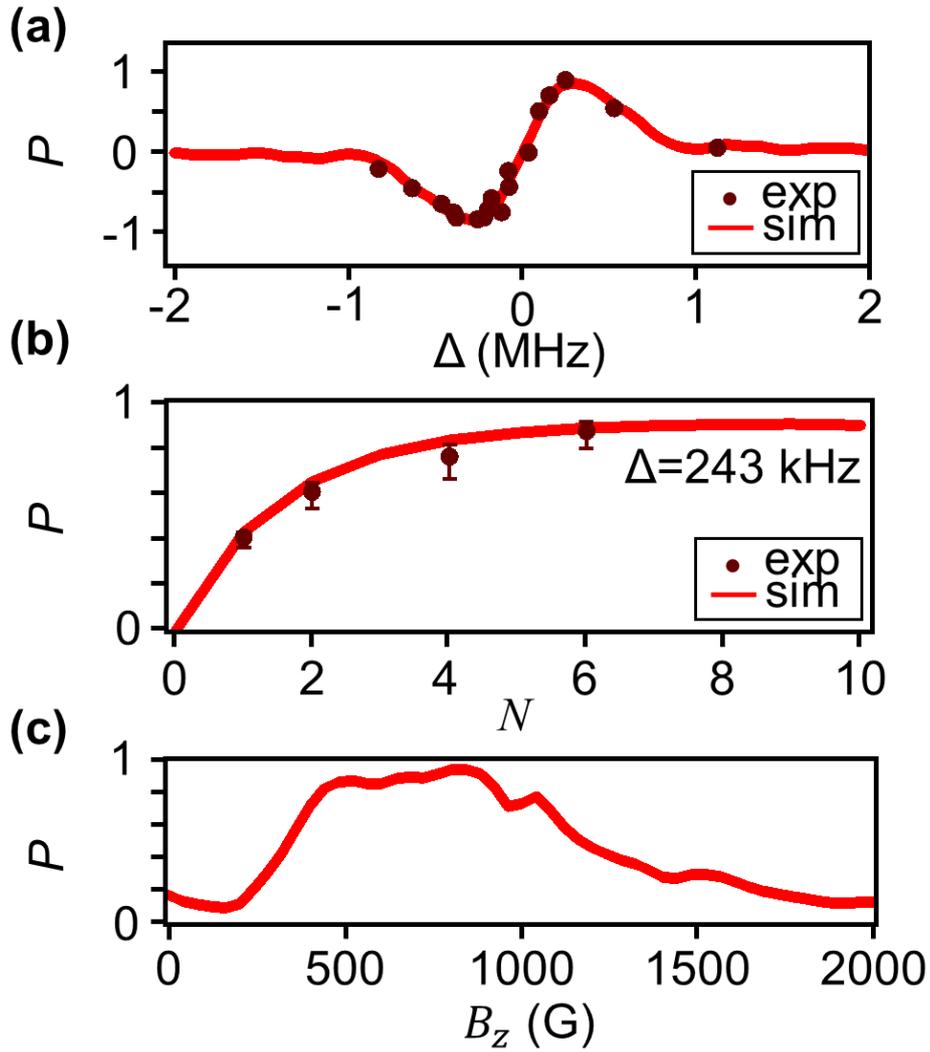

**Figure 3. Numerical simulation. a)** Comparison of experimental (dots) and theoretical (solid curve) nuclear polarization as a function of microwave detuning. **b)** Polarization saturation curve while increasing the number of times the polarization sequence is repeated ($N$). The red line is the simulation result for $A_{zz}$ = 687 kHz and $A_{\text{ani}}$ = 215 kHz; the dark red dots are the experimental results extracted from the time-domain result of the Ramsey sequence. **c)** Maximum polarization calculated while varying the external magnetic field. The polarization efficiency increases as the external magnetic field approaches the condition $A_{zz} = \gamma_c B_z$.

To further support the detailed procedure of the nuclear spin polarization and examine the external magnetic field dependence, we performed time-dependent spin dynamics simulations using the python-based QuTIP package and the Hamiltonian in the rotating frame [22,23]. In order to reflect the experimental conditions, we considered appropriate Lindblad operators providing optically induced relaxation rates; however, we did not consider the dephasing of the NV center during the polarization process as the light is illuminated in a chopped fashion (see Appendix D for details and additional simulation results). In Figure 3, we calculated the experimental polarization based on the time domain result of the Ramsey measurement sequence. We used two sine functions with decay to fit the two oscillations with different frequencies coming from different nuclear spin states, as shown in Fig. 2c. The polarization was calculated and plotted using Eq. (3), where the fitted amplitude of each sine function represents $\mathcal{P}_\uparrow$ and $\mathcal{P}_\downarrow$. Because of the fine time step used for the measurement, the time domain fitting result shows relatively small deviations for the fitting parameters, providing higher precision than the fitting of the FFT result. Figure 3a shows the simulation results of $P$ as a function of $f_{mw}$ at $B_z$ = 520 G. The maximum experimental polarization value obtained from the time domain fitting differs from the values obtained from the FFT result fitting. Using the time domain fitting, we observe that the value of $P$ approaches ~90%(85%) to the $\ket{\uparrow}$ $(\ket{\downarrow})$ orientation. The results are in good agreement with our observations: the qualitative $P$ vs. $f_{mw}$ trend is consistent with the theory and the quantitative maximum $P$ at the resonant conditions matches extremely well with the theoretical value for the given hyperfine parameters obtained by the Ramsey experiment. Figure 3b shows the simulation result of $P$ while varying the polarization sequence number $N$ using the same hyperfine parameters as in Fig. 3a, which is also in good agreement with the theory. Moreover, according to the simulation result for $P$ vs. $B_z$ shown in Fig. 3c, the maximum $P$ occurs when $|\gamma_c B_z|=|A_{zz}|$, which is consistent with our interpretation. We further note that the maximum $P$ occurs in a wide field range around the optimal value, demonstrating the robustness of our method. This robustness comes from nonzero $A_{\text{ani}}$, whose value allows for variations in $B_z$ while maintaining high polarization, provided that the $\theta \sim \pi/2$ condition is met.

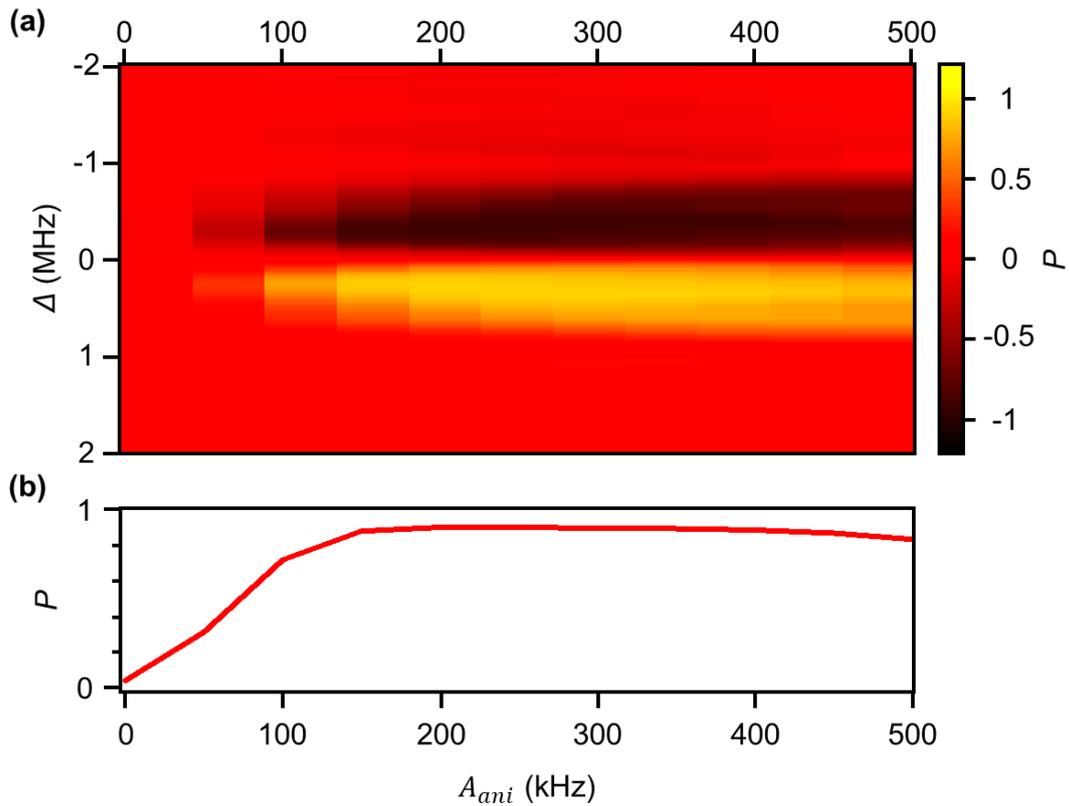

**Figure 4. Estimation of polarization of nuclear spins with varying lattice sites. a)** The nuclear polarization $P$ depending on $A_{ani}$ and polarization microwave detuning. **b)** Trace of maximum polarization value for given $A_{ani}$ extracted from **a)**.

Thus far, we demonstrated efficient spin polarization for a particular $^{13}$C nuclear spin near an NV center. An important question is whether the present protocol can be applied to widely varying locations of $^{13}$C with varying anisotropic tensor components. We turn to discuss this point with the aid of numerical simulation. Figure 4a shows the simulation result of $P$ as a function of $A_{ani}$ and $f_{mw}$ while all other parameters are fixed; and Fig. 4b is the plot of the maximum polarization value among various values of $f_{mw}$ for each given $A_{ani}$ value in Fig. 4a. For a given $A_{zz}$, the dependence of $P$ on $A_{ani}$ reflects the dependence on the relative polar coordinate between the NV center and $^{13}$C resting at different lattice points. For this simulation, $\gamma_c B_z - |A_{zz}|$ is fixed to a small nonzero value, and the result shows that while the resonant condition changes as $A_{ani}$ is varied, near-unity polarization can be achieved for all $A_{ani}$ values as long as the spin-mixing rate is significant set by the condition $|A_{zz}| \sim \gamma_c B_z$, and provided $A_{ani} > \Omega$. Thus, we conclude that the proposed polarization method is not limited to the $^{13}$C investigated in this study but can be generally applied to $^{13}$C spins at widely varying polar angles with respect to the NV direction under conditions that the hyperfine coupling strength of the $^{13}$C satisfies $|A_{zz}| \sim \gamma_c B_z$, and that the selective microwave π pulse with microwave amplitude $\Omega$ satisfies $|A_{zz}| \sim \gamma_c B_z$, $A_{ani} > \Omega$.

**Conclusion**

In conclusion, we have shown that the ground state spin sector of the NV centers coupled with a nearby $^{13}$C nuclear spin has eigenenergy space appropriate for the application of an efficient nuclear spin polarization sequence. Using selective resonant microwave pulses at optimized magnetic field for efficient spin-mixing excitation while using chopped laser pulses providing minimized spin-mixing relaxation, we experimentally demonstrated nuclear polarization, the magnitude and polarity of which can be controlled by the pump frequency. Moreover, we showed that the method is applicable to widely distributed $^{13}$C spins around the NV centers and that for most $^{13}$C spins that are weakly coupled with the NV centers, the maximum polarization can reach > 85% when the nuclear quantization axis is perpendicular to the NV axis. These findings add to the existing nuclear polarization protocols a new method for selectively polarizing relatively weakly coupled nuclear spins in diamond, which is a crucial step toward the initialization of spin registers for quantum information applications. Moreover, as this method provides a means for controlling the environment state from a completely mixed state to pure states, our method can provide a unique opportunity for studying the open quantum system dynamics through controlled system–environment interaction.

**Appendix A: Experimental setup**

**Confocal microscopy**

We used a home-built confocal microscopy setup for excitation and readout of the NV center. A 532-nm diode pumped laser (CNI, MLL-III-532-200 mW) was used to produce the excitation light. This laser goes through an acousto-optic modulator (AOM, Crystal Technology, AOMO 3080-125) for pulsed excitation. The AOM was connected to an 80 MHz driver (Gooch & Housego, 1080AF-DIFO-1.0) providing ~30 ns pulse rise time. A 2-axis galvano mirror (Thorlabs, GVS012) with a telescope system was used for 2D image scanning. The laser was delivered to the sample through an NA 1.3 oil immersion objective (Nikon, CFI Plan Fluor 100X Oil).

A dichroic mirror (Thorlabs, DMLP567) was used to separate excitation and emission beam path. A free space pinhole at the mirror plane of the single mode fiber was used to gain the contrast of the confocal system. The photons are detected using a single photon counter module (avalanche photodiode (APD), Excelitas, SPCM-AQRH-14-FC), which is connected to gated photon counter instruments (Stanford Research Systems, SR400).

**NV photon readout process**

We applied two readout gates during each measurement sequence. The first gate (A gate) was applied at the beginning of the readout laser pulse with a length of 200 ns for spin state readout, and the second gate (B gate) was located 1 μs after the first gate with the same gate length, which was used for the reference. The normalized data, that is the count during A gate / B gate, was used in this study.

**Sample stage setup and magnet alignment**

The diamond sample was mounted on a printed circuit board (PCB) connected to a three-axis piezo stage; the piezo stage was used for position feedback of the NV center, as reported earlier [18], to compensate the slow drift in the position of the NV. An N35 grade neodymium magnet was mounted on the five-axis stage (translation and rotation) for external magnetic field alignment to the NV axis where the alignment was performed by maximizing the photon count. Using this magnet setup, we were able to apply an external magnetic field along the NV center axis ranging from 50 to 1200 G with an alignment precision better than a fraction of a degree.

**Microwave setup**

Two signal generators (Stanford Research Systems SG396, Anritsu MG3700A) were used to generate the microwaves for the Ramsey pulses and polarization sequence. Each signal generator was connected to an arbitrary waveform generator (Tektronix, AFG3252b), which generates the I/Q envelope pulses for the I/Q vector modulation of each microwave. The two microwaves were combined by a power combiner (Hirose, HPS-2C), which then goes through a power amplifier (Mini Circuits, ZHL-42W), and reaches the coplanar waveguide fabricated on the diamond sample. Using this microwave setup, a Rabi frequency up to tens of MHz was achievable. The relative timing of the microwave and green laser pulses were calibrated using a programmable pulse generator (SpinCore Technologies, PulseBlasterESR-PRO 500).

**Temperature tracking**

We implemented a temperature feedback loop to stabilize the temperature of the permanent magnet along with the sample area within ~30 mK fluctuation as shown in Appendix Figure A1. The resonance frequency fluctuation of the NV center due to this temperature fluctuation is within ~50 kHz (N35 temperature coefficient of magnetization $\alpha_0 =$ -0.12% /ºC), which is much smaller than the linewidth of the frequency spectrum of the Ramsey measurement.

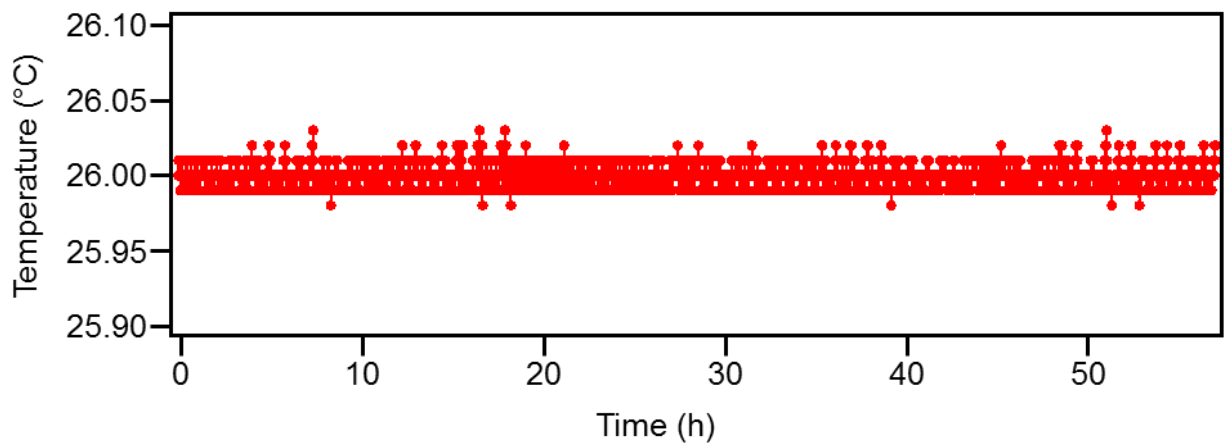

**Appendix Figure A1.** Temperature of the magnet and sample stage area during the measurement process.

**Pulse sequence**

The NV center was initialized to its $m_s = 0$ state by a 5 μs pulsed illumination of the 532-nm green laser. Then, a 2.2 μs rest time was allowed for the relaxation of the NV center electrons existing in the excited or singlet shelving state. After the rest time, the polarization pulse sequence described in the main text was applied: A chopped laser was created using the AOM with 30 ns on / 60 ns off pulse sequence, which was repeated 17 times to complete the initialization of the NV center electron spin. After a 100-ns rest time, a 1.7 μs microwave π pulse was applied, followed by another 100-ns rest time. This polarization sequence comprising chopped laser and microwave application was repeated $N = 6$ times for the saturation of the polarization. Finally, a Ramsey pulse sequence containing 14.5 ns π/2 pulses was used to measure the order of polarization. A phase difference of π/2 was applied for the two π/2 pulses to provide robustness to the gate preparation error. The microwave frequency used in the π/2 pulses was detuned ~5 MHz from the resonant frequency of the $m_s = -1$ manifold, which is smaller than the microwave power used for the pulses. The π gate used in the polarization sequence and π/2 gate for the Ramsey measurement were calibrated by measuring the Rabi oscillation frequency.

**Appendix B: NV center Hamiltonian.**

In this study, the system of interest is a diamond crystal with a nitrogen-vacancy (NV) defect. In the neighborhood of this NV center, there exists $^{13}C$ nuclear spin with 1.1% natural abundance. First, we consider the NV center electron spin and the $^{13}$C nuclear spin independently: The NV center electron spin has a zero-field splitting (*D*) of 2.87 GHz. ($\hbar = 1$):

$$H_{ZF} = DS_z^2 \tag{4}$$

, where $S_z$ is the z component of the spin-1 operator. Under an external magnetic field $B_z$, Zeeman splitting is applied to both the NV center and the $^{13}$C nuclear spin:

$$H_z = \gamma_e S_z B_z + \gamma_c I_z B_z \tag{5}$$

, where the gyromagnetic ratio of the NV center electron (or $^{13}$C nuclear spin) is $\gamma_e \approx 2.8$ MHz/G (or $\gamma_c \approx 1.07$ kHz/G), and $I_z$ is the Pauli spin operator. Next, we consider the dipole interaction of the two spins:

$$H_{dip} = -\frac{\mu_0}{4\pi r^3}\left(\frac{3}{r^2}(\vec{S}\cdot\vec{r})(\vec{I}\cdot\vec{r}) - \vec{S}\cdot\vec{I}\right) \tag{6}$$

Here $\mu_0$ is the vacuum permeability and $\vec{r}$ is the relative position vector between the NV and the $^{13}$C spin. This can be rewritten as:

$$H_{dip} = \vec{S}\cdot\underline{\underline{A}}\cdot\vec{I} \tag{7}$$

where the tensor $\underline{\underline{A}}$ is defined as follows:

$$A_{i,j} = -\frac{\mu_0 \gamma_e \gamma_c}{4\pi r^3}\left(\frac{3}{r^2}r_i r_j - \delta_{i,j}\right), \quad i,j = x,y,z \tag{8}$$

When the external magnetic field is aligned to the NV axis, considering $D - \gamma_e B_z \gg$

$A_{zz}, A_{ani}$, we can apply secular approximation to this system. The Hamiltonian of this system after using secular approximation is expressed as follows [11]:

$$H_0 = DS_z^2 + \gamma_e S_z B_z + \gamma_c I_z B_z + S_z A_{zz} I_z + \frac{A_{ani}}{2} S_z (I_+ e^{-i\phi} + I_- e^{i\phi}) \tag{9}$$

Where is the azimuthal angle between the NV center axis and the position vector between the $^{13}$C nuclear spin and the NV center; and $A_{ani} = \sqrt{A_{zx}^2 + A_{zy}^2}$, where $A_{ij} (i, j = x, y, z)$ is the $ij$ component of the hyperfine tensor; in particular, $A_{zz}$ (or $A_{ani}$) is the position-dependent hyperfine tensor component of $^{13}$C parallel (or perpendicular) to the NV-axis. By diagonalizing this Hamiltonian, we derive the following eigenenergies: $E_{1,2} = \pm \gamma_c B_z / 2$, $E_{3,4} = D - \gamma_e B_z \mp \frac{1}{2}\sqrt{A_{ani}^2 + (A_{zz} - \gamma_c B_z)^2}$, and $E_{5,6} = D - \gamma_e B_z \mp \frac{1}{2}\sqrt{A_{ani}^2 + (A_{zz} + \gamma_c B_z)^2}$, where the corresponding eigenstates expressed in the basis of electron and nuclear spin states $|m_s, m_c\rangle$ are:

$$|\psi_1\rangle = |0, \uparrow\rangle \tag{10}$$

$$|\psi_2\rangle = |0, \downarrow\rangle \tag{11}$$

$$|\psi_3\rangle = \sin(\theta/2) e^{i\phi} |+1, \uparrow\rangle + \cos(\theta/2) |+1, \downarrow\rangle \tag{12}$$

$$|\psi_4\rangle = \cos(\theta/2) |+1, \uparrow\rangle - \sin(\theta/2) e^{-i\phi} |+1, \downarrow\rangle \tag{13}$$

$$|\psi_5\rangle = \cos(\theta'/2) |-1, \uparrow\rangle + \sin(\theta'/2) e^{i\phi} |-1, \downarrow\rangle \tag{14}$$

$$|\psi_6\rangle = -\sin(\theta'/2)e^{-i\phi}|-1,\uparrow\rangle + \cos(\theta'/2)|-1,\downarrow\rangle \quad (15)$$

where $\theta$ and $\theta'$ represent the polar angle between the NV center axis and the position vector between the $^{13}$C nuclear spin and the NV center, which can be calculated as follows:

$$\tan\theta' = \frac{A_{ani}}{-A_{zz} + \gamma_c B_z} \quad (16)$$

$$\tan\theta = \frac{A_{ani}}{A_{zz} + \gamma_c B_z} \quad (17)$$

**Appendix C: Ramsey spectroscopy of the unpolarized spin system.**

By measuring the Ramsey oscillation using microwave π/2 pulses detuned by a few MHz from the resonant frequency of the transition $m_s = 0 \rightarrow -1$, which is expressed as: $\Delta_{(-)} = D - \gamma_e B$, we can measure the energy spectrum of $E_1$, $E_2$, $E_3$, and $E_4$ using the four peak positions in frequency, $p^{(-)}_1$, $p^{(-)}_2$, $p^{(-)}_3$, and $p^{(-)}_4$, derived from its FFT result. The four frequencies that are derived from this measurement correspond to:

$$p^{(-)}_{1,2} = E_{3,4} - E_1 = \Delta_{(-)} \mp \frac{1}{2}\sqrt{A_{ani}^2 + (A_{zz} - \gamma_c B_z)^2} - \gamma_c B_z / 2 \quad (18)$$

$$p^{(-)}_{3,4} = E_{3,4} - E_2 = \Delta_{(-)} \mp \frac{1}{2}\sqrt{A_{ani}^2 + (A_{zz} - \gamma_c B_z)^2} + \gamma_c B_z / 2 \quad (19)$$

However, under conditions where $\tan\theta' = \frac{A_{ani}}{-A_{zz} + \gamma_c B_z} \ll 1$, which can be satisfied when $A_{zz} < 0$, the transitions $|\psi_1\rangle \leftrightarrow |\psi_3\rangle$ and $|\psi_2\rangle \leftrightarrow |\psi_4\rangle$ become forbidden as $\sin\theta \approx 0$, $\cos\theta \approx 1$ in Eq. ( 12 ) and ( 13 ), allowing only $p^{(-)}_1$ and $p^{(-)}_4$ to be visible. Moreover, when $A_{\text{ani}} \ll A_{zz}$, the difference between the two peaks is equal to $A_{zz}$, which is measured to be ~600 kHz, as shown in in Fig. 1e in the main text.

Similarly, for the $m_s = +1$ manifold, using microwaves detuned from $\Delta_{(+)} = D + \gamma_e B_z$ for pulses, the energy–peak conversion relation is expressed as follows:

$$p^{(+)}_{1,2} = E_{5,6} - E_1 = \Delta_{(+)} \mp \frac{1}{2}\sqrt{A_{\text{ani}}^2 + (A_{zz} + \gamma_c B_z)^2} - \gamma_c B_z / 2 \qquad (20)$$

$$p^{(+)}_{3,4} = E_{5,6} - E_2 = \Delta_{(+)} \mp \frac{1}{2}\sqrt{A_{\text{ani}}^2 + (A_{zz} + \gamma_c B_z)^2} + \gamma_c B_z / 2 \qquad (21)$$

Under the conditions: $A_{zz} + \gamma_c B_z \sim 0$, $A_{\text{ani}} \ll A_{zz}$, $p^{(+)}_1$ and $p^{(+)}_2$ (or $p^{(+)}_3$ and $p^{(+)}_4$) become degenerate, and only the following two frequencies can be observed during the Ramsey oscillation:

$$p^{(+)'}_1 = \Delta_{(+)} - \gamma_c B_z / 2 \qquad (22)$$

$$p^{(+)'}_2 = \Delta_{(+)} + \gamma_c B_z / 2 \qquad (23)$$

Therefore, from the difference of the two frequencies, we can measure $\gamma_c B_z$, which is ~600 kHz, as shown in Fig. 1d of the main text. This value is consistent with the magnetic field extracted from the frequency of the two peaks measured in the ESR measurement, which is 520 G. In this study, we used an external magnetic field of 520 G and considered $A_{zz} \sim$ -600 kHz.

In this study, we used the $m_s = +1$ manifold for the polarization sequence, and the $m_s = -1$ manifold was used for Ramsey sequence time-domain measurement to avoid the strong nuclear spin mixing that occurs at the $m_s = +1$ manifold. The different manifolds show different frequency behaviors, as shown in Eq. ( 18 ), ( 19 ), ( 22 ), and ( 23 ).

For the $m_s = +1$ state, the oscillation frequency originating from the $m_c = \uparrow$ state is lower than the frequency from $m_c = \downarrow$. On the other hand, for the $m_s = -1$ state, the oscillation frequency originating from the $m_c = \uparrow$ state is higher than the frequency from $m_c = \downarrow$, explaining the reversed frequency position of the dominant peak shown in Fig. 2d and 2e in the main text.

**Appendix D: Details on the numerical simulation.**

**Calculation of the degree of polarization $P$**

Using the rotating wave approximation, the rotating frame Hamiltonian becomes:

$$H_{rot} = H_0 + w_{mw} S_z + \Omega S_x \tag{24}$$

, where $H_0$ is defined in Eq. ( 9 ), $w_{mw}$ is the applied microwave frequency, and $\Omega$ is the Rabi amplitude. For the calculation, we used the following master equation in the Lindblad form:

$$\frac{d\rho}{dt} = -i[H_{rot}, \rho] + \sum (L\rho L^\dagger - \frac{1}{2}\{L^\dagger L, \rho\}) \tag{25}$$

This equation contains the following Lindblad operators:

$$L^{gl}_{up,\uparrow(\downarrow)} = \sqrt{\Gamma_{gl} n_{th}} \left| -1, \uparrow(\downarrow) \right\rangle \left\langle 0, \uparrow(\downarrow) \right| \quad (26)$$

$$L^{gl}_{down,\uparrow(\downarrow)} = \sqrt{\Gamma_{gl}(1+n_{th})} \left| 0, \uparrow(\downarrow) \right\rangle \left\langle -1, \uparrow(\downarrow) \right| \quad (27)$$

$$L^{D}_{i} = \sqrt{\Gamma^{D}_{i}} \left| \psi_i \right\rangle \left\langle \psi_i \right|, \quad i=1,2,3,4 \quad (28)$$

$$L^{gl}_{12(21)} = \sqrt{\Gamma^{gl}_{n}} \left| \psi_{1(2)} \right\rangle \left\langle \psi_{2(1)} \right| \quad (29)$$

Each operator represents the decay rate caused by the green laser illumination ($\Gamma_{gl}$), transverse relaxation ($\Gamma^{D}_{i}$), and optically induced transverse relaxation ($\Gamma^{gl}_{12(21)}$) [21]. In order to capture realistic experimental results, we included in the simulation the time-dependent pulse sequence that was used to polarize our $^{13}$C nuclear spin. By solving the Lindblad master equation using this Hamiltonian in Eq. ( 24 ), we calculated the variation in the polarization result as a function of the various environmental factors of the NV center, such as the external magnetic field strength and the hyperfine coupling strength, as shown in Figs. 3 and 4 in the main text.

We calculated the solution of the Eq. ( 25 ) using the built-in *mesolve* function of the QuTIP (Quantum toolbox in Python) package. During the calculation, we considered the full time dependence of the pulse sequence (for example, switching on and off of Ω and $\Gamma_{gl}$) reflecting the experimental condition.

The initial density matrix was set as a completely mixed state of $|\psi_1\rangle = |0,\uparrow\rangle$ and $|\psi_2\rangle = |0,\downarrow\rangle$, which is expressed as: $\rho_0 = |0\rangle_e \langle 0|_e \otimes \frac{1}{2}(|\uparrow\rangle_c \langle \uparrow|_c + |\downarrow\rangle_c \langle \downarrow|_c)$. To calculate the order of polarization after the completion of the polarization process, we calculated

$P = \dfrac{\mathcal{P}_1 - \mathcal{P}_2}{\mathcal{P}_1 + \mathcal{P}_2}$, where $\mathcal{P}_1 = \langle \psi_1 | \rho(t_f) | \psi_1 \rangle$ and $\mathcal{P}_2 = \langle \psi_2 | \rho(t_f) | \psi_2 \rangle$. The calculated probability of each nuclear spin $\mathcal{P}_{1(2)}$ is different from the probability $\mathcal{P}_{\uparrow(\downarrow)}$ defined in the main text, but becomes equal at the end of each single polarization step where the state of the NV center spin is completely initialized to $m_s = 0$, which are the points where we can experimentally measure the order of polarization.

**Fitting $P$ vs $f_{mw}$**

The fitting of Fig. 3a in the main text was conducted using the *curve_fit* function built in SciPy. The fitting function was defined to solve the master equation containing pulse sequences with green laser time $t_{gl}$ and microwave $\pi$ pulse time $t_{mw}$. This pulse sequence was repeated $N = 6$ times; and $t_{gl}$, $t_{mw}$, and $N$ match the condition used for the experiment. The fitting parameters for this function are $f_{rel}$, $A_{zz}$, and $A_{ani}$, where $f_{rel}$ is defined to match the center frequency $\Delta = 0$ between the experiment and simulation, and $A_{zz}$ and $A_{ani}$ are the hyperfine interaction strengths. The initial guess for the parameters of the *curve_fit* function was defined as: $f_{rel} = 0$, $A_{zz} = 600$ kHz, and $A_{ani} = 100$ kHz. The result of the *curve_fit* function is shown in Fig. 3a, where the resulting fitting parameters are: $A_{zz} = 687$ kHz and $A_{ani} = 215$ kHz, which are close to the experimental values. The subsequent simulations share the hyperfine coefficients obtained from this process. Following is a summary of the additional simulation data and the corresponding parameter tables:

**Appendix Table A1.** Table of coefficients used for each simulation

| Coefficients | Fitting $P$ vs $f_{mw}$ | $P$ vs $N$ | $P$ vs $B_z$ | $P$ vs $B_z$ and $A_{ani}$ |
|---|---|---|---|---|
| $t_{gl}$ | 300 ns | 300 ns | 300 ns | 300 ns |
| $t_{mw}$ | 1.7 µs | 1.7 µs | 1.7 µs | 20 µs |
| $\Omega$ | 294.1176 kHz | 294.1176 kHz | 294.1176 kHz | 25 kHz |
| $\Gamma_{gl}$ | 8 MHz | 8 MHz | 8 MHz | 8 MHz |
| $D$ | 2.87 GHz | 2.87 GHz | 2.87 GHz | 2.87 GHz |
| $\gamma_e$ | 2.8 MHz/G | 2.8 MHz/G | 2.8 MHz/G | 2.8 MHz/G |
| $\gamma_c$ | 1.07 kHz/G | 1.07 kHz/G | 1.07 kHz/G | 1.07 kHz/G |
| $\Gamma_i^D$ | 0 | 0 | 0 | 0 |
| $\Gamma_{12(21)}^{gl}$ | 0 | 0 | 0 | 0 |
| $n_{th}$ | 0 | 0 | 0 | 0 |
| $A_{zz}$ | - | 686.5546 kHz | 686.5546 kHz | 625 kHz |
| $A_{ani}$ | - | 215.3535 kHz | 215.3535 kHz | - |
| $B_z$ | 520 | 520 | - | 520 |
| $N$ | 6 | - | 6 | 10 |

*P* vs *N*

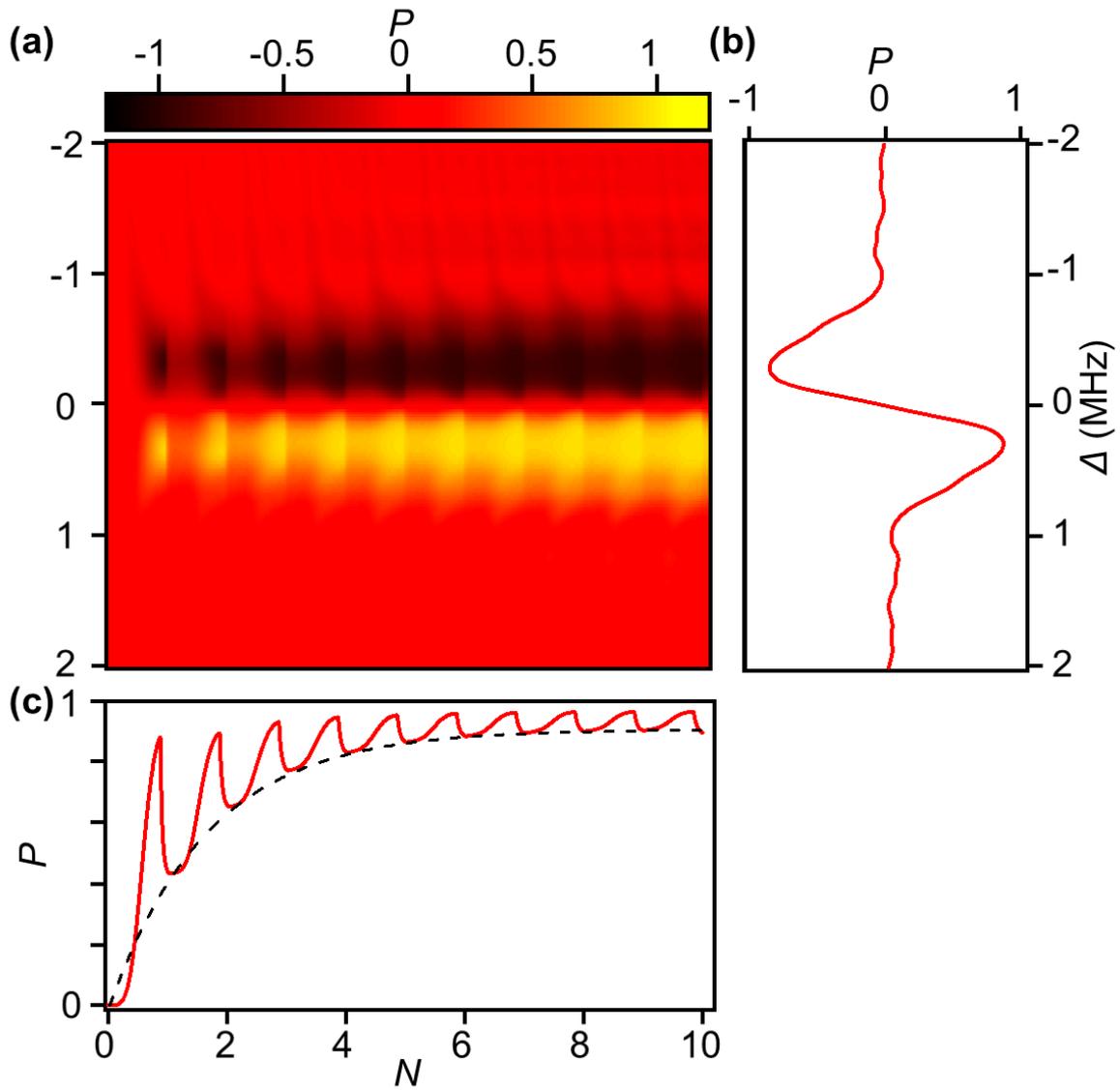

**Appendix Figure A2. Numerical simulation of *P* vs *f*<sub>mw</sub> and *t*(*N*) a)** Simulation result of *P* as a function of $\Delta$ and time. The evolution of the states was calculated and plotted with 10 ns steps along the x axis during the whole polarization sequence. The x axis is shared with **c)**, where N represents the time when the $N^{th}$ polarization step ends. The y axis is shared with **b)** showing the detuning of each line. **b)** Line cut along *N* = 6 showing the dependence of polarization on detuning frequency. This line cut is similar to the simulation result displayed in Fig. 3a of the main text. **c)** Line cut along $\Delta$ = 290 kHz showing *P* vs *N*. The points on the line cut with integer *N* values are used as the simulation result in Fig. 3b of the main text. The black dashed line is a guide to the eye for the exponential saturation of the actual polarization achieved after the integer polarization sequences. The oscillating feature shown in **a)** and **c)** is a feature that comes from the repetition of the pulse sequence. The increasing part of the oscillation is a feature of the state evolution while applying the selective microwave π pulse and the decreasing part is a feature of the state evolution while applying the green laser.

*P* vs *B*$_z$

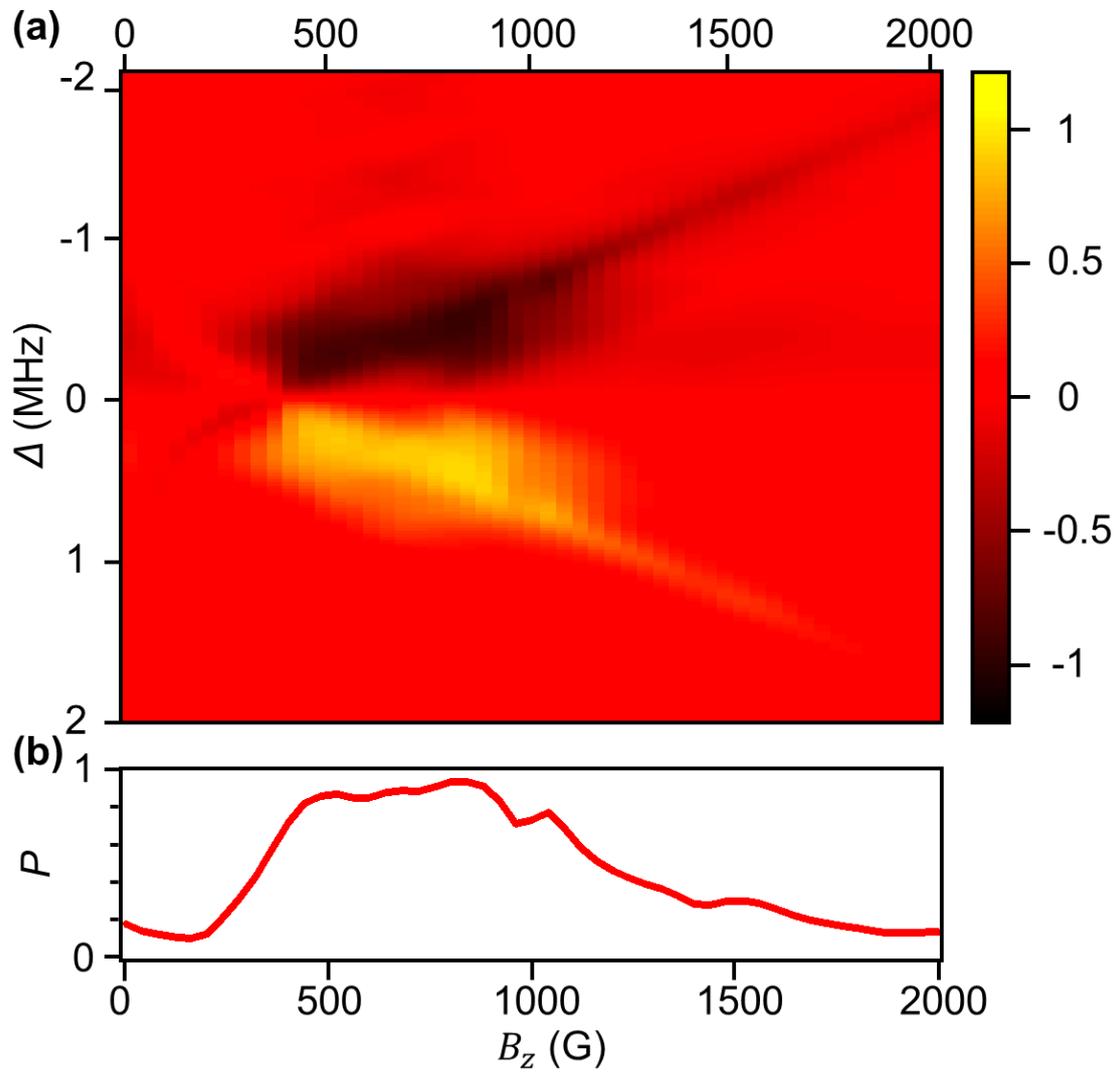

**Appendix Figure A3. Numerical simulation of *P* vs *f*$_{mw}$ and *B*$_z$** **a)** Simulation of *P* as a function of $\Delta$ and *B*$_z$. **b)** Maximum polarization value for each line with given *B*$_z$ in **a)**.

*P* vs *B*$_z$ and *A*$_{ani}$

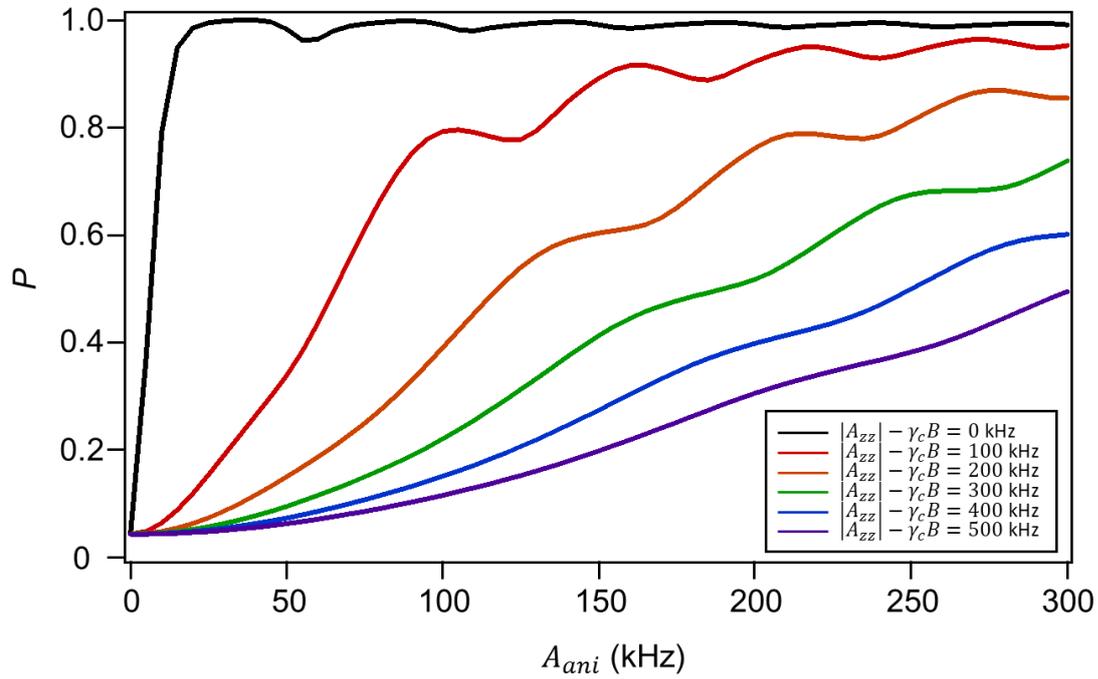

**Appendix Figure A4. Numerical simulation of maximum *P* vs *A*$_{ani}$ and *B*$_z$** Simulation of maximum polarization while varying *B*$_z$ and *A*$_{ani}$. The result shows that the polarization becomes more robust to *B*$_z$ as the value of *A*$_{ani}$ increases.

**Acknowledgements**

This work was supported by the National Research Foundation of Korea (NRF) Grant funded by the Korean Government (MSIT) (No. 2015R1A5A1037668 and No. 2018R1A2A3075438) and the Creative-Pioneering Researchers Program through Seoul National University (SNU).